# Title

Comprehensive determination of Burgers vectors of threading dislocations in GaN substrates by combining reflection and transmission synchrotron-radiation x-ray topography

# Authors

Kazuki Ohnishi[1, a)], Kenji Iso[2], Hirotaka Ikeda[2], Yoshiyuki Tsusaka[3], and Yongzhao Yao[1]

[a)] Corresponding author: E-mail: ohnishi@icsdf.mie-u.ac.jp

# Affiliations

*[1]Innovation Center for Semiconductor and Digital Future, Mie University, 1577, Kurimamachiya-cho, Tsu 514-8507, Japan*

*[2]Mitsubishi Chemical Corporation, Ushiku, Ibaraki 300-1295, Japan*

*[3]Graduate School of Science, University of Hyogo, 3-2-1, Koto, Kamigori-cho, Ako-gun, Hyogo 678-1297, Japan*

## Abstract

Burgers vectors (***b***) of threading dislocations (TDs) in an acidic ammonothermal-grown GaN substrate were investigated using synchrotron radiation x-ray topography (SR-XRT) by combining both reflection and transmission modes. Reflection XRT images recorded with six equivalent ***g*** vectors of $11\overline{2}4$ revealed spot-like contrasts corresponding to TDs. Based on the contrast conditions, the possible Burgers vectors were constrained, and the *c*-axis component of ***b*** for mixed-type TDs was estimated from the contrast size. Using transmission XRT images recorded under several two-beam diffraction conditions, the (0001) in-plane direction of ***b*** was evaluated based on the ***g·b*** invisibility criterion. Furthermore, by analyzing the linewidths of dislocation images observed under kinematical diffraction contrast, the magnitude of the a-axis component of ***b*** was determined. By combining these analyses, the Burgers vectors of individual TDs, including edge- and mixed-type dislocations, were determined. In addition, a pair of screw-type TDs with opposite Burgers vectors, $\pm 1c$, was observed in the transmission SR-XRT. These results demonstrate that the combined use of reflection and transmission SR-XRT provides a practical approach for complete determination of Burgers vectors in GaN substrates.

## 1. Introduction

Gallium nitride (GaN), which is a major wide-bandgap semiconductor, is one of the key materials in an energy-saving and low-carbon-emitting society owing to its superior properties such as its high breakdown electric field, high electron mobility, and high saturation velocity [1-5]. Many reports have demonstrated that vertical GaN power devices, such as metal-oxide-semiconductor transistors and p-n junction diodes, have high potential for power devices [2, 6-16]. It is well known that dislocations have negative impacts on device performance. Especially, some dislocations cause leakage current and the degradation of reliability, and such dislocations are known as killer defects. So far, screw-type threading dislocation (TD) with the Burgers vector ($\boldsymbol{b}$) of 1$\boldsymbol{c}$ has been reported as one of the killer defects [17-20]. However, dislocations other than screw-type TDs with $\boldsymbol{b}$ = <0001> may also act as killer defects, and their identification is therefore important for improving the performance and reliability of GaN power devices. To further improve device performance, reliability, and yield, it is crucial to identify killer defects and to elucidate their distribution within GaN substrates.

There are several methods to investigate dislocations: x-ray topography (XRT) [21-32], etch pit method [33], transmission electron microscopy [17, 19, 34], cathodoluminescence [35], multiphoton excitation photoluminescence [36], spontaneous and stimulated Raman spectral mapping [37, 38], birefringence method [39] and phase-contrast microscopy [40]. Among these methods, XRT is a powerful tool for characterizing dislocations because dislocations and their $\boldsymbol{b}$ can be identified nondestructively over a wide area. Also, the dynamics of defects during the device operation or annealing can be observed by in-situ operand XRT [41, 42]. Additionally, the three-dimensional observation of dislocation propagations can be recorded [43, 44].

In general, XRT has two geometries: the reflection mode (Bragg case) and the transmission mode (Laue case). The reflection mode has been widely used for investigating dislocations in GaN substrates since TDs near the surface can be observed in a wide range. On the other hand, the transmission mode has rarely been applied to the observation of dislocations in GaN substrates, although it enables observation of dislocation propagation inside the substrates. This is because heavy Ga atoms result in strong x-ray absorption, which makes transmission XRT difficult under a kinematical diffraction phenomenon. To mitigate x-ray absorption, the transmission XRT has been performed using a GaN substrate thinned to about 50 μm [28]. However, this approach cannot be applied to the dislocation characterization in GaN substrates with a typical thickness of 350 μm. Recently, the transmission XRT was demonstrated for characterization of dislocations in thick GaN substrates by utilizing the anomalous transmission of x-ray [29, 31, 32]. This anomalous transmission, which is known as the Borrmann effect, occurs under a dynamical x-ray diffraction phenomenon [45, 46]. Yao *et al*. reported that the synchrotron-radiation (SR-) XRT observation of dislocation propagation in GaN substrates using the super-Borrmann effect under the multi-beam diffraction condition, and this approach is powerful for the characterization of dislocation in thick substrates [32]. In this way, the transmission mode is expected to become increasingly important for the characterization of dislocations in thick GaN substrates.

Despite the advantages of these XRT configurations, several challenges remain for the accurate determination of Burgers vectors of dislocations in GaN substrates. In the reflection mode, the size of the spot-contrast depends on $|\boldsymbol{b}|$, while the shape of the contrast reflects the direction of $\boldsymbol{b}$. However, the spatial resolution of x-ray films is not sufficient to determine $\boldsymbol{b}$ accurately from the observed contrasts [26]. In addition, reflection XRT can observe only dislocations located near the surface of the substrate. In the transmission mode based on the Borrmann effect, dislocation contrasts are formed under dynamical diffraction conditions, which makes the interpretation of the contrasts more complicated. Although previous studies have reported the observation of edge-type TDs with an *a*-axis component [32], experimental data for other types of dislocations remain limited. Therefore, the interpretation of such contrasts requires both the development of simulation techniques based on dynamical diffraction theory and the accumulation of experimental data. Owing to these limitations, it is difficult to accurately estimate $\boldsymbol{b}$ using either method alone. Therefore, in this study, we demonstrate that the combined use of reflection and transmission SR-XRT enables comprehensive determination of Burgers vectors of individual TDs in thick GaN substrates.

## 2. Experimental procedure

An n-type GaN (0001) substrate with a TD density on the order of $10^3$ $\mathrm{cm}^{-2}$ fabricated by the acidic ammonothermal growth method was used [47]. The off-cut angle was about 0.4 ° toward the *m*-axis. Both sides of the substrate were subjected to mirror-polished surfaces to perform the SR-XRT without surface damage. The reflection SR-XRT was performed at beamline BL-3C of the Photon Factory (2.5 GeV storage ring) at the High Energy Accelerator Research Organization (KEK), Japan. This beamline is the same as that used in Ref. 23. Figure 1(a) shows the schematic illustration of the experimental setup in the reflection geometry. The grazing-incidence mode was applied to the (0001) surface at incident angles of about 5 °. To distinguish the *a*-axis components of ***b***, the six-equivalent ***g*** vectors of $11\bar{2}4$ were chosen. The incident x-ray wavelength was set to 1.40 Å using a double crystal Si (111) monochromator and the beam size was about 19 mm × 6 mm. Topographic images were recorded on x-ray films (Agfa Structurix D2), which were placed about 15 cm from the sample surface, perpendicular to the diffracted x-ray.

The transmission SR-XRT was performed at the beamline BL24XU of SPring-8 (8 GeV storage ring), Japan [48]. Since the x-ray beam at this beamline can be regarded as the plane wave, the dynamical x-ray diffraction phenomena can be observed in GaN substrates. Figure 1(b) shows the schematic illustration of experimental setup in the transmission geometry for the six-beam diffraction condition. The incident x-ray wavelength was set to 1.24 Å, and the beam size was 1.2 × 1.2 $mm^2$. Note that this wavelength corresponds to the low-energy side of Ga K-edge (1.20 Å) and a $\mu t$ value is 5.4 ($\mu$: linear absorption coefficient, $t$: crystal thickness), indicating that the normal transmission intensity without the Borrmann effect is reduced to about 1% of that under $\mu t$ of 1. At an incident angle of 27 °, the six-beam diffraction condition, consisting of the forward-transmitted wave (o-wave) and five diffracted waves with ***g*** = $01\bar{1}0$, $\bar{1}2\bar{1}0$, $\bar{2}200$, $\bar{2}110$, and $\bar{1}010$, can be achieved by adjusting $\varphi$, $\psi$, and $\omega$ angles shown in Fig. 1(b). The wave vectors of $\boldsymbol{K_0}$–$\boldsymbol{K_5}$ corresponds to the o-wave and five diffracted waves with ***g*** = $01\bar{1}0$, $\bar{1}2\bar{1}0$, $\bar{2}200$, $\bar{2}110$, and $\bar{1}010$, which are hereafter denoted as $\boldsymbol{g_1}$, $\boldsymbol{g_2}$, $\boldsymbol{g_3}$, $\boldsymbol{g_4}$, and $\boldsymbol{g_5}$, respectively. This is because GaN has a hexagonal crystal structure with the space group of $P6_3mc$. After adjusting the six-beam diffraction condition, the $\psi$ angle was rotated by 0.005 ° to obtain the two-beam diffraction condition for each ***g*** vector, and XRT images were recorded under each ***g*** vector by rotating the $\omega$ angle in steps of 0.0002 °. Details of the setup for the six-beam diffraction condition and the five two-beam diffraction conditions are described in Ref. 32. In addition to the above conditions, XRT images were also recorded under the two-beam diffraction condition for ***g*** = $\bar{1}100$ denoted as $\boldsymbol{g_6}$ because $\boldsymbol{g_6}$ provides a similar structure factor with $\boldsymbol{g_1}$ and $\boldsymbol{g_5}$ and is more suitable than $\boldsymbol{g_3}$ for comparing XRT images recorded with $\boldsymbol{g_1}$ and $\boldsymbol{g_5}$. In this configuration, the incident angle changed to about 12.7 ° maintaining the incident x-ray wavelength of 1.24 Å. XRT images were recorded using the o-wave. To record the o-wave, the imaging system consisting of a scintillator, relay lenses, and a CMOS detector was used. The spatial resolution of the imaging system was 0.65 μm/pixel [30].

## 3. Results and discussion

Figure 2 shows the reflection XRT images of the same area taken with the six-equivalent ***g*** vectors of $11\bar{2}4$. Spot-shaped bright or dark contrasts are seen in the image, corresponding to threading dislocations. In this area, the threading dislocation density was about $7 \times 10^3$ $cm^{-2}$. Here, we focus on TDs labeled with E1, E2, E3, M1, and M2. These dislocations exhibit either bright or dark contrast depending on the ***g*** vector. In particular, bright spots are observed in four images, whereas dark spots are observed in the other two images. Ray-tracing simulations based on the lattice misorientation caused by dislocation-induced displacement, together with the experimental results, indicate that edge- and mixed-type TDs can appear as either bright or dark spots, whereas screw-type TDs appear only as bright spots. This behavior originates from the direction of the *a*-axis component of ***b*** with respect to the ***g*** vector [23, 25, 26, 49]. Based on the contrast conditions for the six ***g*** vectors, the possible Burgers vectors of TDs labeled with E1, E2, E3, M1, and M2 were constrained as summarized in Table 1. In particular, these ***b*** were identified to be edge- or mixed-type with ***b*** of an *a*-axis component. Although rigorous estimation based on the spot size is difficult due to the insufficient spatial resolution of the x-ray films, the *c*-axis component of ***b*** can be reasonably estimated to be ±1***c*** if these TDs are mixed-type. This is because the lattice constant along the *c*-direction (5.19 Å) is longer than that along the *a*-direction (3.19 Å); thus, the contribution of the *c*-axis component to the magnitude of ***b*** is larger than that of the *a*-axis component in mixed-type dislocations. If the *c*-axis component were ±2***c***, the resulting lattice misorientation would be expected to produce contrast spots larger than those observed. In contrast, such a comparison cannot be readily applied to edge-type TDs. Some edge-type TDs have relatively large |***b***|, and therefore their spot contrasts can be comparable in size to those of mixed-type TDs with ***b*** of $\langle 11\bar{2}0\rangle/3 + \langle 0001\rangle$. In fact, TDs with the large *a*-axis component have been reported in GaN substrates [27, 32].

To further investigate characteristics of TDs, transmission XRT was performed. Five strong Laue spots were confirmed on the fluorescent screen, corresponding to the diffracted beam of $\boldsymbol{g}_1$–$\boldsymbol{g}_5$, as shown in Fig. 3(a). The transmission XRT images were obtained with the o-wave under this six-beam diffraction, and TDs were clearly observed as shown in Fig. 3(b). This indicates that GaN substrates fabricated by the acidic ammonothermal growth method have a high crystalline quality because the Borrmann effect occurs in crystals with a high degree of perfection. Especially, the six-beam condition, i.e. super-Borrmann effect, requires a higher degree of perfection than the two-beam condition. Therefore, the acidic ammonothermal growth is a promising method for fabricating GaN substrates with a high crystalline quality. Using transmission XRT images recorded under five two-beam diffraction conditions for $\boldsymbol{g}_1$–$\boldsymbol{g}_5$, $\boldsymbol{b}$ of TDs was evaluated based on the $\boldsymbol{g}\cdot\boldsymbol{b}$ invisibility criterion.

Figure 4 shows the $\omega$ scan of the integral intensity of the o-wave at $\psi$ = 0.005 °. Transmission XRT images recorded with the o-wave at the exact Bragg condition of $\boldsymbol{g}_1$–$\boldsymbol{g}_5$ are also presented. It should be noted that the trend of the integral intensity for $\boldsymbol{g}_1$–$\boldsymbol{g}_5$ arises from differences in the structure factors of $01\bar{1}0$, $\bar{1}2\bar{1}0$, $\bar{2}200$ families, as well as from the varying strength of the Borrmann effect for different sets of atomic planes. It is seen that some TDs were nearly invisible depending on the diffraction condition. For more detailed analysis, transmission XRT images recorded with the o-wave at the exact Bragg condition of $\boldsymbol{g}_1$, $\boldsymbol{g}_5$, and $\boldsymbol{g}_6$ are shown in Fig. 5. The XRT image recorded with $\boldsymbol{g}_6$ was used for the $\boldsymbol{g}\cdot\boldsymbol{b}$ invisibility criterion instead of $\boldsymbol{g}_3$, because the structure factor was chosen to be identical to those of $\boldsymbol{g}_1$ and $\boldsymbol{g}_5$, allowing reliable comparison of the XRT images. It is seen that TDs with E1, E2, E3 and M1 are clearly visible in Figs. 5(b) and (c), but nearly invisible in Fig. 5(a). This indicates that those TDs have $\boldsymbol{b}$ which are perpendicular to $\boldsymbol{g}_1$. In other words, $\boldsymbol{b}$ of those TDs has an $a$-axis component parallel or antiparallel to $[2\bar{1}\bar{1}0]/3$. Thus, TDs labeled E1, E2, and E3 were determined to be edge-type, whereas TDs labeled M1 were determined to be mixed-type with a $c$-axis component of $[000\bar{1}]$. Note that the magnitude of the $a$-axis component of $\boldsymbol{b}$ cannot be determined from the $\boldsymbol{g}\cdot\boldsymbol{b}$ invisibility criterion. On the other hand, the TDs labeled M2 were clearly visible in Figs. 5(a), (b), and (c), despite being identified by reflection SR-XRT as having Burgers vectors with an $a$-axis component.

From the above results, $\boldsymbol{b}$ of TDs labeled M2 could not be estimated and the magnitude of the $a$-axis component of $\boldsymbol{b}$ for the TDs labeled E1, E2, E3, and M1 could not be determined. To solve these issues, we focused on dislocation lines exhibiting kinematical diffraction contrast. Such contrast can be obtained under a large deviation from the exact Bragg condition ($\Delta\omega$) [32, 50]. Under this condition, only strongly distorted regions near the dislocation core satisfy the Bragg condition, whereas perfect crystal regions do not. As a result, diffraction mainly occurs in the highly strained region near the dislocation core. In this region, the wavefield cannot fully adapt to the rapidly varying lattice distortion. Consequently, the diffraction behavior approaches the kinematical regime, and the dislocation contrast appears as straight and thin lines. According to Suvorov *et al.*[51], dislocation contrasts under the two-beam diffraction condition is primarily attributed to interbranch scattering. Near the dislocation core, enhanced interbranch scattering allows Bloch waves belonging to one sheet of the dispersion surface to generate components corresponding to the other sheet. The strongly absorbed component decays within the crystal, thereby reducing the intensity of the wavefield reaching the exit surface and weakening the dynamical diffraction contrast. Consequently, the dynamical diffraction contrast is significantly reduced, and the observed contrast becomes closer to kinematical behavior. Under kinematical diffraction contrast, the scattered area is proportional to $|\boldsymbol{g} \cdot \boldsymbol{b}|$ [52]. Accordingly, the apparent linewidth of the dislocation image is proportional to $\sqrt{|\boldsymbol{g} \cdot \boldsymbol{b}|}$. Previously, our group analyzed the linewidths of edge-type TD images in the GaN substrate and distinguished their ***b*** with magnitude of 1***a*** and 2***a*** [32]. In this paper, ***b*** of TDs labeled M2 and the magnitude of the *a*-axis component of ***b*** for TDs were determined by analyzing the linewidth of each dislocation image.

The linewidth of the TD labeled E1 image was analyzed first. Figure 6(a) shows the transmission XRT image of the TD labeled E1 recorded with the o-wave at $\boldsymbol{g}_5$ under $\Delta\omega$ for −0.003 °. In this condition, the dislocation line was clearly observed under the kinematical diffraction. The linewidth of the TD labeled E1 image was determined by fitting with a Gaussian profile, as shown in Fig. 6(b). The full width at half maximum (FWHM) of the dislocation intensity profile was determined to be 3.8 μm. This value is in good agreement with that for an edge-type TD image with ***b*** = 1***a*** reported in our previous paper [32]. ***b*** of TDs labeled E1 is therefore determined to be $[2\bar{1}\bar{1}0]/3$. Subsequently, the linewidths of the other dislocation images were analyzed in the same manner, and their FWHMs were compared with that of the TD labeled E1 image, whose $|\boldsymbol{b}|$ had already been determined. Figures 6(c) and (d) show the FWHM ratio of the TDs labeled E1, E2, E3, M1, and M2 relative to the TD labeled E1. To determine ***b*** of TDs labeled M2, the analysis was performed for transmission XRT images recorded with $\boldsymbol{g}_4$ and $\boldsymbol{g}_5$, respectively. For TDs labeled E2 and M1, the FWHM ratios relative to the TD labeled E1 are comparable for $|\boldsymbol{g} \cdot \boldsymbol{b}| = 1$ for both $\boldsymbol{g}_4$ and $\boldsymbol{g}_5$. Accordingly, ***b*** of TDs labeled E2 and M1 were determined to be $[\bar{2}110]/3$ and $[2\bar{1}\bar{1}\bar{3}]/3$, respectively. The FWHM ratio of the TD labeled E3 is comparable for $|\boldsymbol{g} \cdot \boldsymbol{b}| = 4$ and 2 for $\boldsymbol{g}_4$ and $\boldsymbol{g}_5$, respectively. This result indicates that ***b*** = $[\bar{4}220]/3$. That is, the TD labeled E3 has ***b*** = 2***a***, with its direction along $[\bar{2}110]/3$. TDs with the *a*-axis component of 2***a*** have been reported in ammonothermal grown GaN substrates [27, 32]. The FWHM ratio of the TD labeled M2 is comparable for $|\boldsymbol{g} \cdot \boldsymbol{b}| = 1$ for both $\boldsymbol{g}_4$ and $\boldsymbol{g}_5$, respectively. Based on this result, ***b*** of TDs labeled M2 can be estimated as follows. Reflection and transmission XRT images revealed that TDs labeled M2 are edge- or mixed-type TDs. If they were edge-type TDs, the linewidth for $\boldsymbol{g}_5$ would not be observed due to the $\boldsymbol{g}\cdot\boldsymbol{b}$ invisibility criterion.

Hence, they are identified as mixed-type TDs. To determine $\boldsymbol{b}$ of mixed-type TDs labeled M2 the linewidth for $\boldsymbol{g}_4$ is examined. $\boldsymbol{b}$ of these TDs are expected to be $n[\bar{2}110]/3 + [0001]$ or $n[11\bar{2}0]/3 + [000\bar{1}]$ (where $n$ is the natural number). Among these candidates, only $[11\bar{2}0]/3 + [000\bar{1}]$ satisfies $|\boldsymbol{g} \cdot \boldsymbol{b}| = 1$ for $\boldsymbol{g}_4$. Therefore, $\boldsymbol{b}$ of TDs labeled M2 is estimated to be $[11\bar{2}\bar{3}]/3$. The procedure used to estimate $\boldsymbol{b}$ of the labeled TDs is summarized in Fig. 7.

Based on the above results, the contrasts from the TDs labeled M2 should disappear in the transmission XRT images recorded with $\boldsymbol{g}_3$ and $\boldsymbol{g}_6$. However, these contrasts are faintly observed, as discussed above. This behavior is attributed to the condition $\boldsymbol{g}\cdot(\boldsymbol{b} \times \boldsymbol{l}) \neq 0$ ($\boldsymbol{l}$ denotes the line direction of the dislocation) even when $\boldsymbol{g}\cdot\boldsymbol{b} = 0$, as well as to the surface relaxation effect at the surface termination points [53].

Finally, the results of reflection and transmission XRT for screw-type TDs are discussed. A pair of screw-type TDs with opposite Burgers vector directions was observed in the reflection XRT recorded with $\boldsymbol{g} = 11\bar{2}4$, as shown in Fig. 8(a). According to the simulation, the separation distance between the two dislocations was about 5 μm, and their Burgers vectors were $\pm 1\boldsymbol{c}$. For SiC crystals, the nucleation of such pairs of screw-type TDs has been reported to occur at inclusions [54]. Similar pairs of screw-type TDs have also been observed in GaN substrates grown by basic ammonothermal growth [26]. These observations suggest that the nucleation mechanism of such pairs may be common to both basic and acidic ammonothermal growth. Figures 8(b)-(f) show the transmission XRT images of the pair of screw-type TDs recorded with the o-wave at $\boldsymbol{g}_1$–$\boldsymbol{g}_5$ under two-beam diffraction conditions. Spot-like contrasts were observed in each XRT image, whereas the contrasts related to the dislocation line were absent. The absence of the dislocation line and the appearance of spot-like contrasts were due to the satisfaction of $\boldsymbol{g}\cdot\boldsymbol{b} = 0$ and the surface relaxation effect, respectively, as mentioned above. Therefore, screw-type TDs can be identified by observing only spot-like contrasts without dislocation-line contrasts for $\boldsymbol{g}_1$–$\boldsymbol{g}_5$. Based on the displacement filed of the surface relaxation effect (Eshelby twist) for screw-type TDs [53], the size of spot-like contrasts is expected to depend on $\boldsymbol{b}$, and $|\boldsymbol{b}|$ of screw-type TDs may be identified by analyzing the spot-like contrasts. Further investigation will be carried out to evaluate the relationship between the spot-like contrast size and $|\boldsymbol{b}|$ for screw-type TDs.

## 4. Conclusion

Burgers vectors of TDs in an acidic ammonothermal-grown GaN substrate were systematically estimated using SR-XRT in both reflection and transmission modes.

Reflection SR-XRT recorded with six equivalent $\boldsymbol{g}$ vectors of $11\bar{2}4$, enabled classification of threading dislocations based on their contrast conditions and allowed the *c*-axis component of $\boldsymbol{b}$ for mixed-type TDs to be estimated from the contrast size. Transmission SR-XRT utilizing anomalous transmission recorded with $\boldsymbol{g}_1$–$\boldsymbol{g}_5$ further provided information on the (0001) in-plane direction of $\boldsymbol{b}$ through the $\boldsymbol{g}\cdot\boldsymbol{b}$ invisibility criterion. In addition, analysis of the linewidth of dislocation images under kinematical diffraction contrast enabled determination of the magnitude of the *a*-axis component of $\boldsymbol{b}$. By combining these analyses, the Burgers vectors of edge- and mixed-type TDs were unambiguously determined.

In addition, a pair of screw-type threading dislocations with opposite $\boldsymbol{b}$ observed in the reflection SR-XRT image. Transmission SR-XRT images of this pair of screw-type TDs exhibited only spot-like contrasts without dislocation-line contrasts, which is attributed to the satisfaction of $\boldsymbol{g}\cdot\boldsymbol{b} = 0$ and the surface relaxation effect at the dislocation termination points. These present results demonstrate that the combination of reflection and transmission SR-XRT enables reliable comprehensive determination of both the direction and magnitude of Burgers vectors of TDs in GaN substrates.

**Acknowledgment**

This research was supported by JSPS KAKENHI Grant Number 23K13572. The synchrotron XRT observations were performed at Photon Factory, KEK under proposal No. 2024G520, and at SPring-8 with approval from the Japan Radiation Research Institute under proposal No. 2025R3304. We would like to thank Professor Keiichi Hirano with KEK for his support in maintaining the BL-3C beamline at the Photon Factory.

**Author Contributions**

**Kazuki Ohnishi**: Conceptualization (lead), Formal analysis (lead), Investigation (lead), Methodology (equal), Visualization (lead), Writing-original draft (lead), Writing – review & editing (equal).

**Kenji Iso**: Resources (equal), Investigation (equal), Writing – review & editing (equal).

**Hirotaka Ikeda**: Resources (equal), Investigation (equal), Writing – review & editing (equal).

**Yoshiyuki Tsusaka**: Resources (equal), Investigation (equal), Writing – review & editing (equal).

**Yongzhao Yao**: Conceptualization (equal), Formal analysis (equal), Investigation (equal), Methodology (equal), Resources (equal), Writing – review & editing (equal), Project

Administration (lead)

**Data availability**

The data that support the findings of this study are available from the corresponding author upon reasonable request.

# Figures and captions

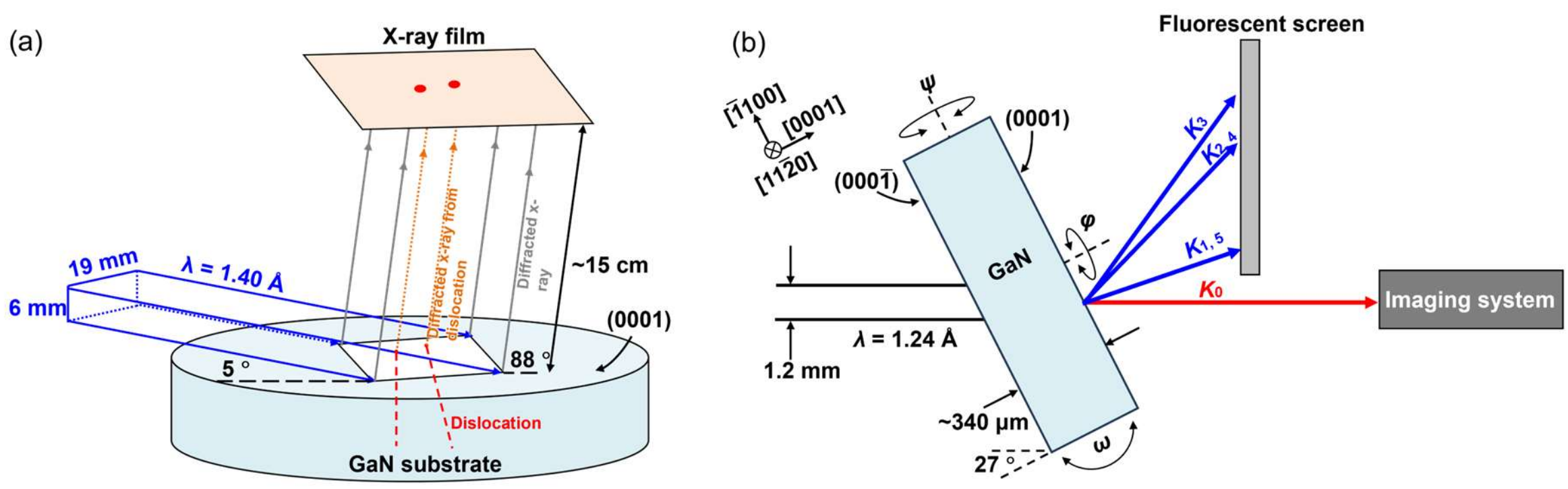

**Fig. 1.** Experimental setup for SR-XRT with (a) the reflection and (b) the transmission modes. In the transmission mode, the six-beam diffraction condition was illustrated.

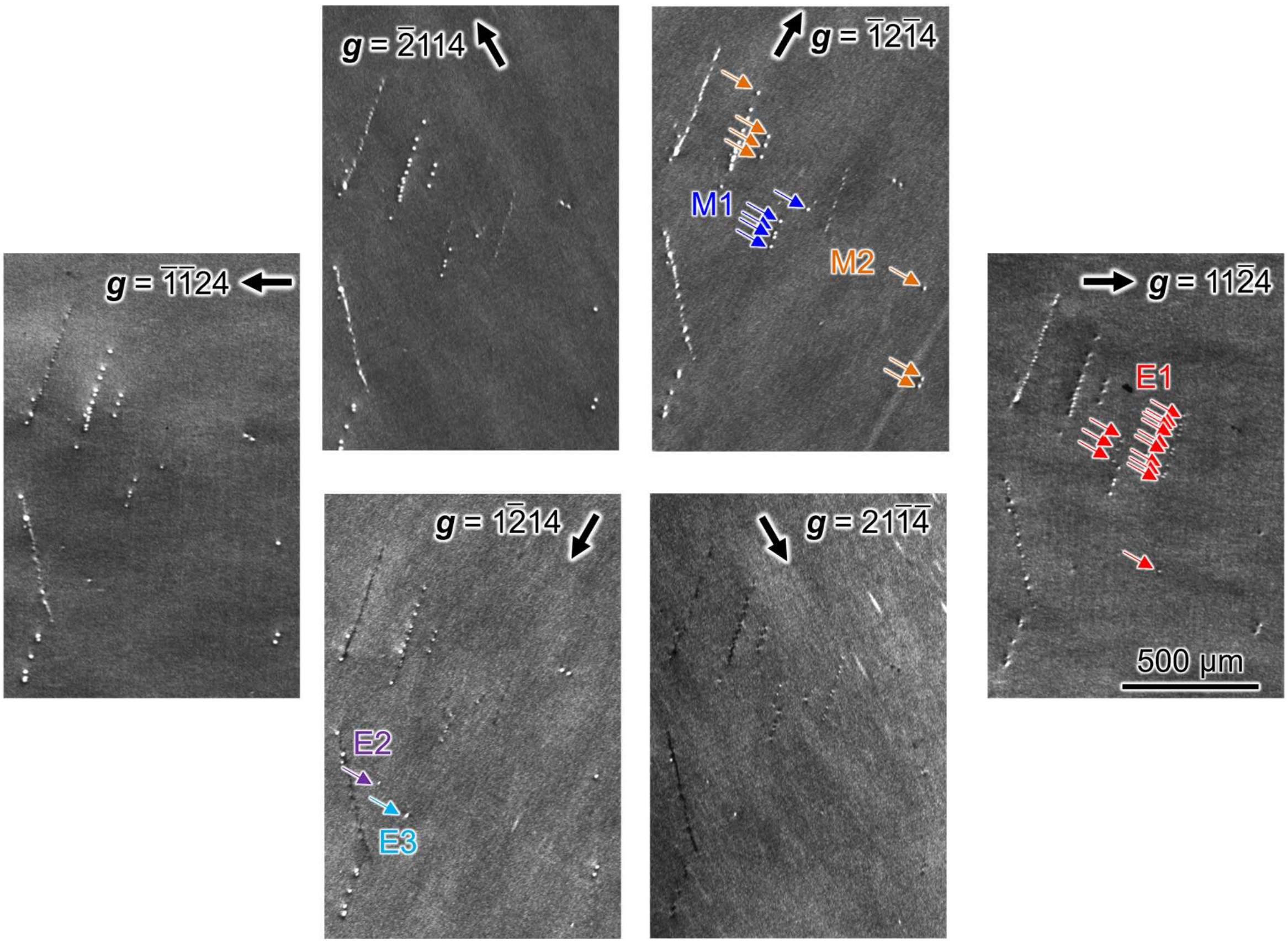

**Fig. 2.** Reflection XRT images of the same area taken with the six-equivalent ***g*** vectors of $11\bar{2}4$.

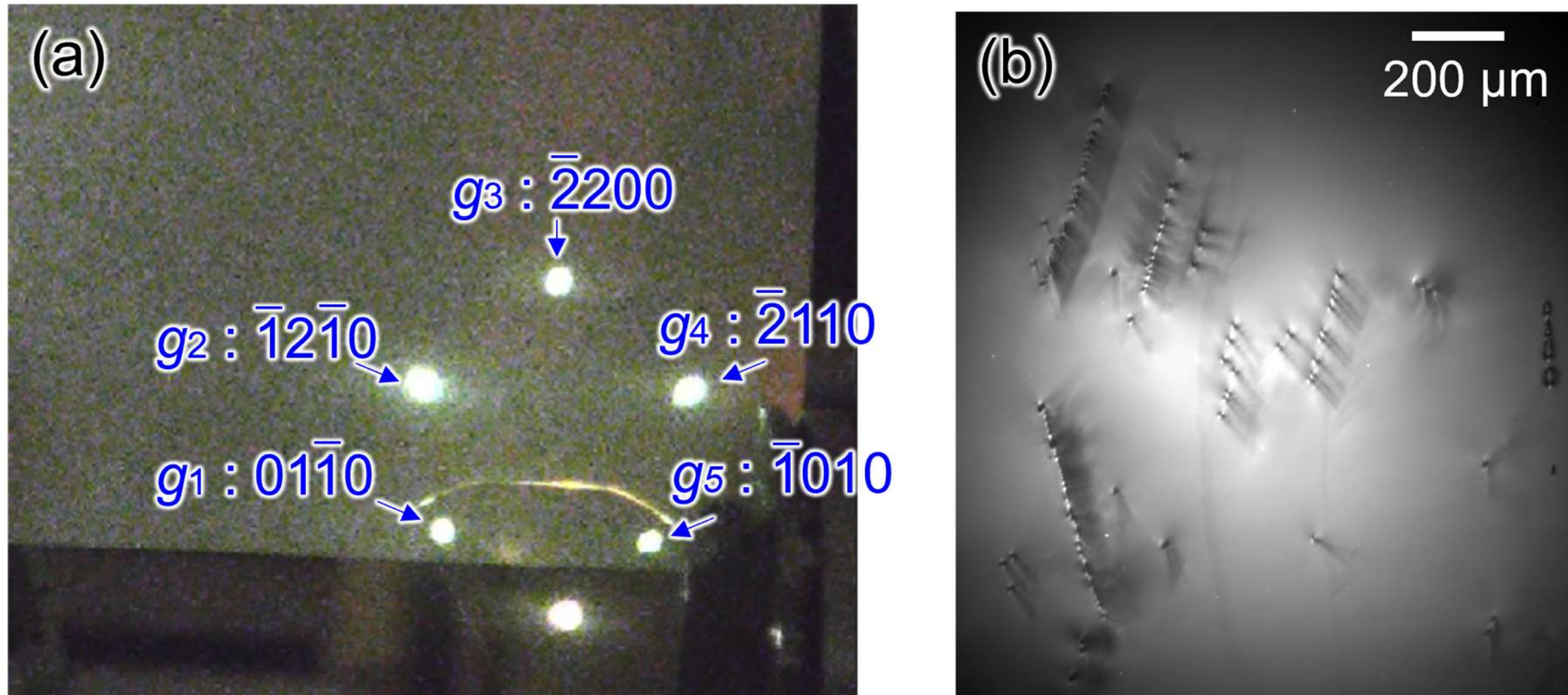

**Fig. 3.** (a) Photograph of the five Laue spots of diffraction beam of $\boldsymbol{g}_1$–$\boldsymbol{g}_5$, and (b) transmission XRT image recorded with the o-wave under the six-beam diffraction condition.

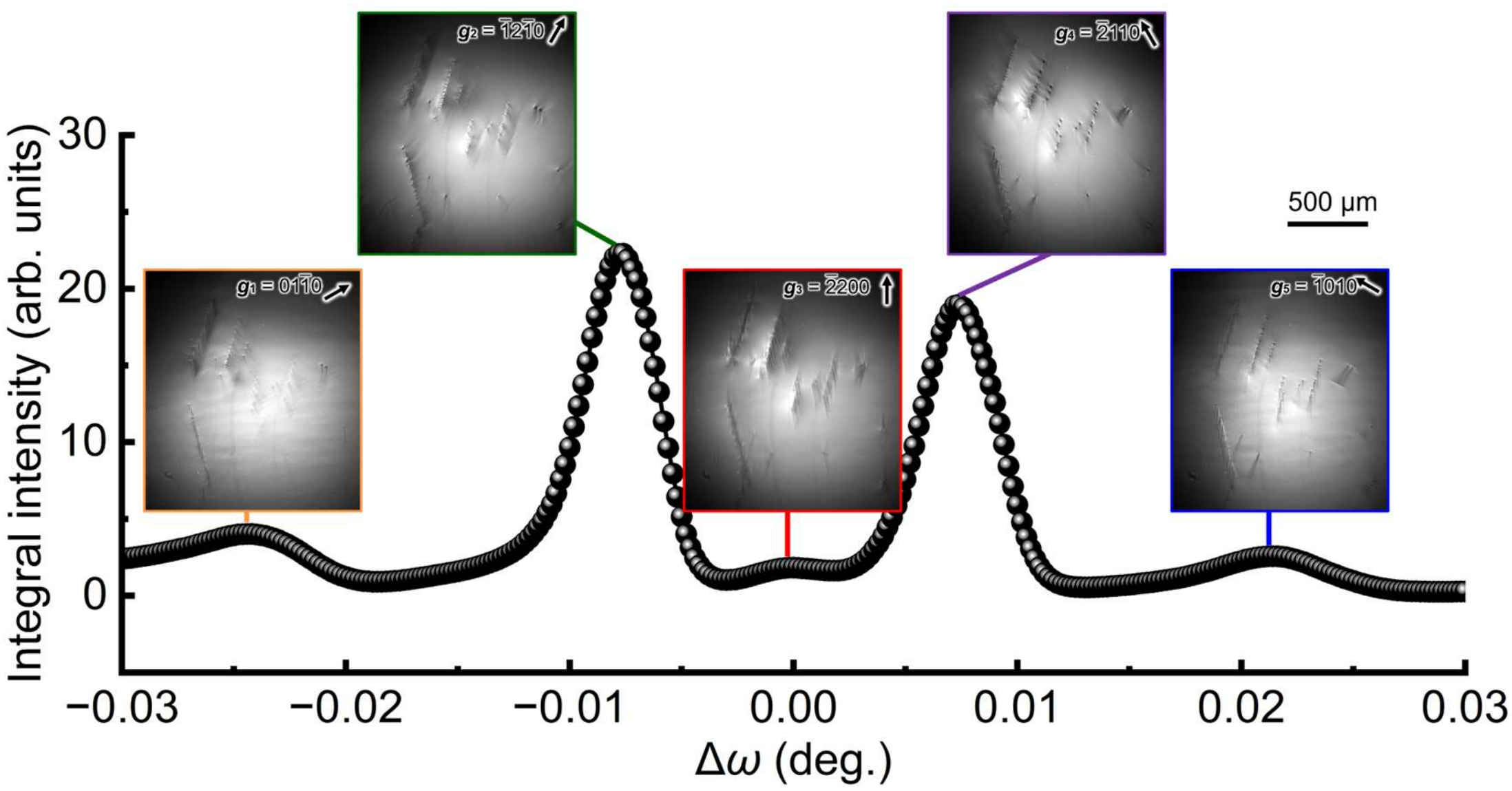

**Fig. 4.** $\omega$ scan of the integral intensity of the o-wave at $\psi = 0.005$ °. Transmission XRT images recorded with the o-wave at the exact Bragg condition of $\boldsymbol{g}_1$–$\boldsymbol{g}_5$ are also presented.

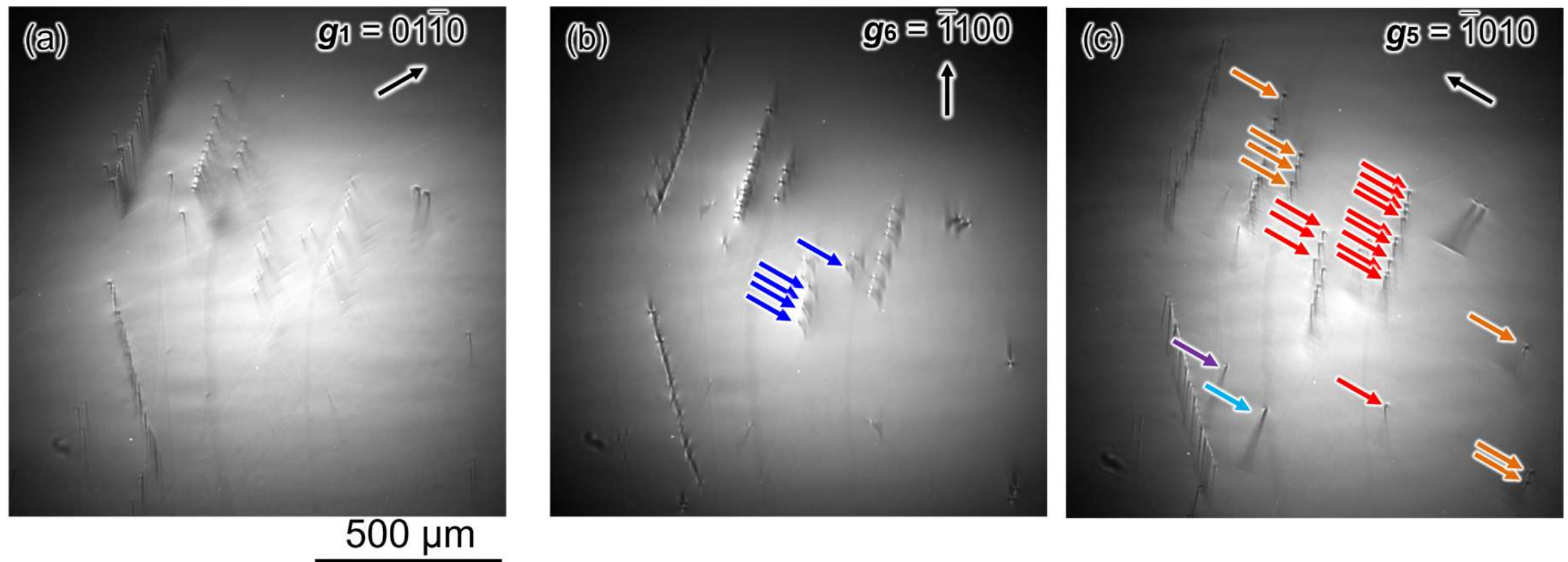

**Fig. 5.** Transmission XRT images recorded with the o-wave at the exact Bragg condition of (a) $\boldsymbol{g}_1$, (b) $\boldsymbol{g}_6$, and (c) $\boldsymbol{g}_5$.

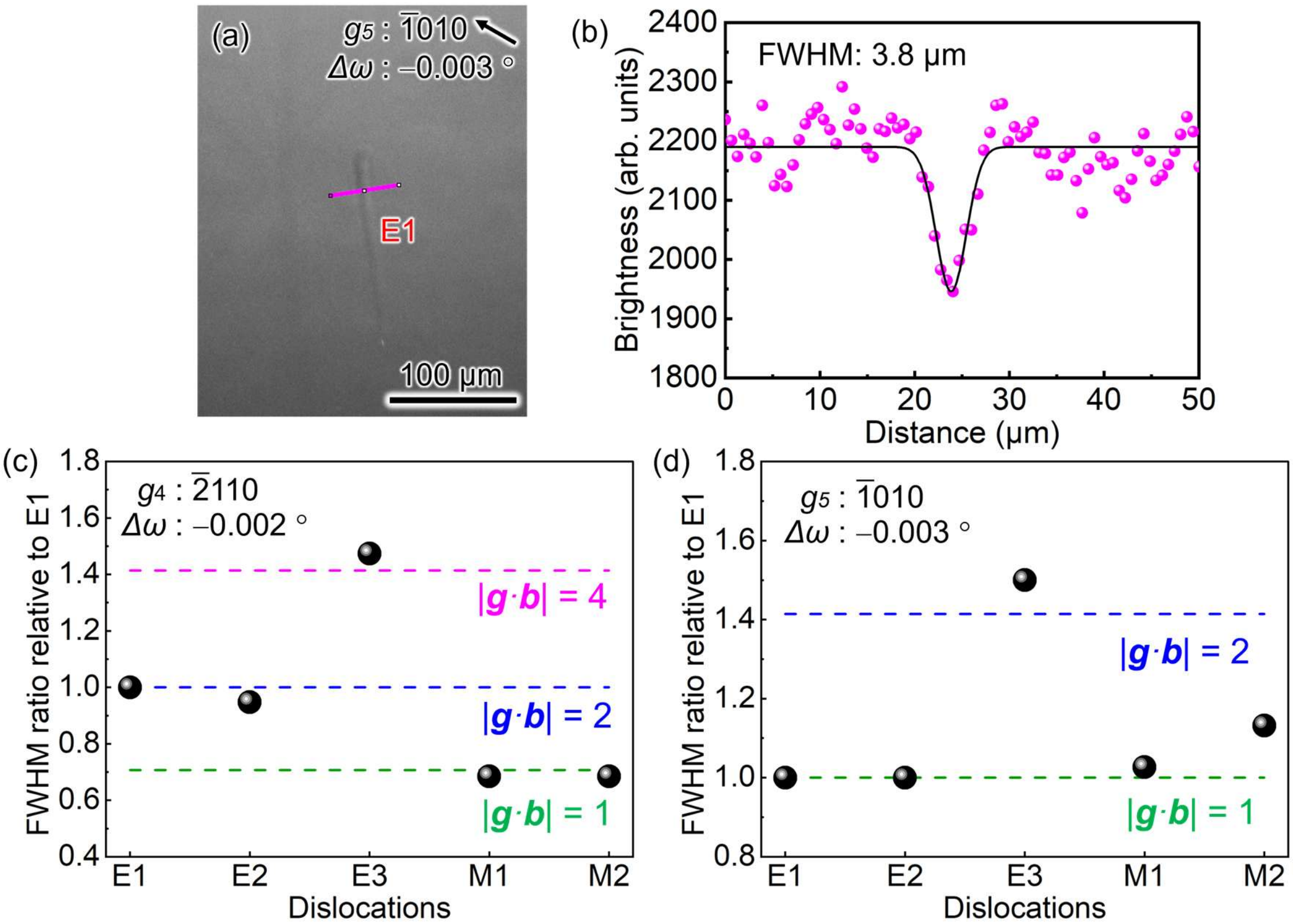

**Fig. 6.** (a) Transmission XRT image of the TD labeled E1 recorded with the o-wave at $\boldsymbol{g}_5$ under $\Delta\omega$ for −0.003 °. (b) The linewidth of the TD image labeled E1under the kinematical diffraction condition. Dot shows the Pixel brightness profiles, and the black line shows the Gaussian fitting profile. the FWHM ratio of the TDs relative to the TD labeled E1 obtained from the transmission XRT images recorded with (c) $\boldsymbol{g}_4$ and (d) $\boldsymbol{g}_5$ under the kinematical diffraction condition.

1. **Reflection SR-XRT recorded with six equivalent *g* vectors of $11\bar{2}4$**
   - **Contrast conditions: Classification of TD type**
   - **Contrast size: Magnitude *c*-axis component of *b* for screw- and mixed-type TDs**

| ***b*** | Edge-type TD | Mixed-type TD | |
|---|---|---|---|
| E1 | $n[2\bar{1}\bar{1}0]/3$ | $n[11\bar{2}0]/3 + [0001]$ | $n[1\bar{2}10]/3 + [000\bar{1}]$ |
| E2 | $n[\bar{2}110]/3$ | $n[\bar{1}\bar{1}20]/3 + [0001]$ | $n[\bar{1}2\bar{1}0]/3 + [000\bar{1}]$ |
| E3 | $n[\bar{2}110]/3$ | $n[\bar{1}\bar{1}20]/3 + [0001]$ | $n[\bar{1}2\bar{1}0]/3 + [000\bar{1}]$ |
| M1 | $n[11\bar{2}0]/3$ | $n[\bar{1}2\bar{1}0]/3 + [0001]$ | $n[2\bar{1}\bar{1}0]/3 + [000\bar{1}]$ |
| M2 | $n[\bar{1}2\bar{1}0]/3$ | $n[\bar{2}110]/3 + [0001]$ | $n[11\bar{2}0]/3 + [000\bar{1}]$ |

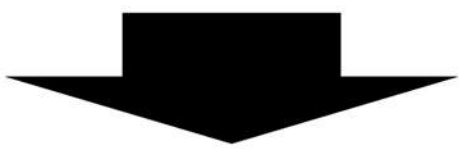

2. **Transmission SR-XRT recorded with $g_1$−$g_6$**
   - ***g·b* invisibility criterion: Constrains possible (0001) in-plane directions of *b***

| ***b*** | Edge-type TD | Mixed-type TD | |
|---|---|---|---|
| E1 | $n[2\bar{1}\bar{1}0]/3$ | $n[11\bar{2}0]/3 + [0001]$ | $n[1\bar{2}10]/3 + [000\bar{1}]$ |
| E2 | $n[\bar{2}110]/3$ | $n[\bar{1}\bar{1}20]/3 + [0001]$ | $n[\bar{1}2\bar{1}0]/3 + [000\bar{1}]$ |
| E3 | $n[\bar{2}110]/3$ | $n[\bar{1}\bar{1}20]/3 + [0001]$ | $n[\bar{1}2\bar{1}0]/3 + [000\bar{1}]$ |
| M1 | $n[11\bar{2}0]/3$ | $n[\bar{1}2\bar{1}0]/3 + [0001]$ | $n[2\bar{1}\bar{1}0]/3 + [000\bar{1}]$ |
| M2 | $n[\bar{1}2\bar{1}0]/3$ | $n[\bar{2}110]/3 + [0001]$ | $n[11\bar{2}0]/3 + [000\bar{1}]$ |

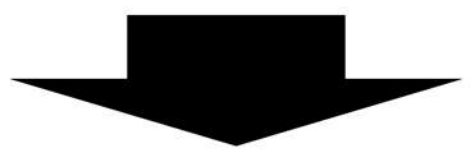

3. **Kinematical diffraction contrasts of transmission SR-XRT image**
   - **Analysis linewidth of dislocation image (Linewidth $\propto \sqrt{|g \cdot b|}$): Determination of magnitude of *a*-axis component of *b* and remaining in-plane direction of *b***

| ***b*** | Edge-type TD | Mixed-type TD | |
|---|---|---|---|
| E1 | $[2\bar{1}\bar{1}0]/3$ | $n[11\bar{2}0]/3 + [0001]$ | $n[1\bar{2}10]/3 + [000\bar{1}]$ |
| E2 | $[\bar{2}110]/3$ | $n[\bar{1}\bar{1}20]/3 + [0001]$ | $n[\bar{1}2\bar{1}0]/3 + [000\bar{1}]$ |
| E3 | $[\bar{4}220]/3$ | $n[\bar{1}\bar{1}20]/3 + [0001]$ | $n[\bar{1}2\bar{1}0]/3 + [000\bar{1}]$ |
| M1 | $n[11\bar{2}0]/3$ | $n[\bar{1}2\bar{1}0]/3 + [0001]$ | $[2\bar{1}\bar{1}0]/3 + [000\bar{1}]$ |
| M2 | $n[\bar{1}2\bar{1}0]/3$ | $n[\bar{2}110]/3 + [0001]$ | $[11\bar{2}0]/3 + [000\bar{1}]$ |

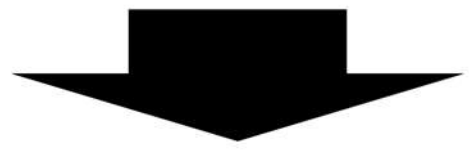

**Determination of Burgers vectors of TDs**

**Fig. 7.** Procedure for determining the Burgers vectors of TDs using combined reflection and transmission SR-XRT. Reflection SR-XRT constrains the possible Burgers vectors and provides information on the *c*-axis component. Transmission SR-XRT further constrains the possible (0001) in-plane direction of ***b*** using the ***g***·***b*** invisibility criterion. The remaining ambiguity in (0001) the in-plane direction and the magnitude of the *a*-axis component are determined by linewidth analysis under kinematical diffraction contrast of transmission XRT images.

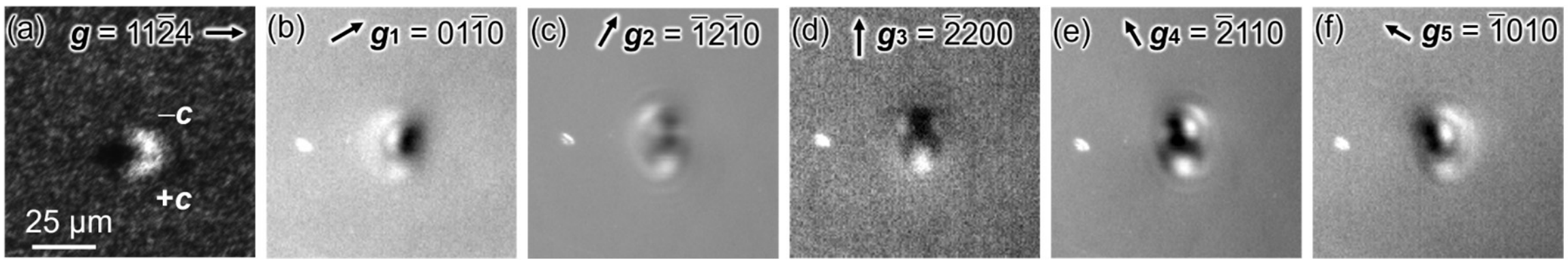

**Fig. 8.** The pair of the screw-type TDs with opposite ***b*** of ±[0001] observed by (a) the reflection SR-XRT recorded with $\boldsymbol{g} = 11\bar{2}4$ and the transmission SR-XRT recorded with (b)-(f) $\boldsymbol{g}_1$–$\boldsymbol{g}_5$.